\documentclass[aps,prl,twocolumn]{revtex4-2}
\usepackage{amssymb}
\usepackage{amsmath}
\usepackage{epsfig}
\usepackage{float}
\usepackage{hyperref}
\begin{document}
\title{A General Scenario of Tunneling Time in Different Energy Regimes}
\author{Sheng-Chang Li}
\email{scli@xjtu.edu.cn}

\affiliation{MOE Key Laboratory for Nonequilibrium Synthesis and Modulation of Condensed Matter, Shaanxi Province Key Laboratory of Quantum Information and Quantum Optoelectronic Devices, and School of Physics, Xi'an Jiaotong University, Xi'an 710049, China}

\begin{abstract}
We theoretically study the tunneling time by investigating a wave packet of Bose-condensed atoms passing through a square barrier. We find that the tunneling time exhibits different scaling laws in different energy regimes. For negative incident energy of the wave packet, counterintuitively, the tunneling time decreases very rapidly with decreasing incident velocity. In contrast, for positive incident energy smaller than the barrier height, the tunneling time increases slowly and then reaches a maximum, which is in agreement with the Larmor clock experiments. The effect of the barrier width related to the uncertainty principle on the maximum tunneling time is also addressed. Our work provides a general scenario of tunneling time that can be used to understand and explain the controversy over tunneling time.
\end{abstract}

\maketitle

Tunneling is an important quantum mechanical effect, which describes one microcosmic particle having a finite probability of crossing an energy barrier when the particle's energy is less than the barrier. Continuous research on quantum tunneling stems from both the incomplete understanding of the tunneling process itself and its extensive applications in many fields. The tunneling time describing the specific process of tunneling events, which is closely related to the uncertainty principle, has always been controversial. The first attempt to calculate tunneling time can be traced back to MacColl's work in 1932 \cite{prl1}. Since then, great efforts have been made to define \cite{prl2,prl3,prl4} and measure \cite{prl5,prl6,prl7,prl8,prl9} tunneling times. %Theoretical research is mainly focused on whether the concept of tunneling time is superluminal or violates causality \cite{nat16}. 
%Two theoretical concepts of the tunneling time --- ``phase time" or ``Wigner time" \cite{nat2} and ``interaction time" or ``semiclassical time" \cite{nat14,nat15} were proposed. These two kinds of time have been measured experimentally and qualitatively agreed well with theoretical predictions \cite{nat3,nat4,nat6,nat20}.
Recently, based on the attoclock \cite{prl11}, the tunneling time defined as the time that the tunneling electron of an atom spends under a barrier formed by a strong laser field and atomic Coulomb potential has been detected in strong-field ionization experiments \cite{prl10,nat23,nat24,nat25}. However, there are controversies over theoretical explanations for the experimental results \cite{nat28,prl2016,nat7,nat24,cyj}. One reason is that in this case, it is difficult to clearly define the tunneling time due to the time-dependent laser-Coulomb-formed barrier. Another reason is that such attosecond-scale tunneling time is deduced from experimental observable (i.e., photoelectron momentum distribution) via strong-field theory models and cannot be measured directly.  

On the other hand, ultracold atoms, especially Bose--Einstein condensates (BECs), as macroscopic quantum matter, provide unprecedented opportunities and ideal platforms for the experimental detection of tunneling time. In the most recent study \cite{nat2020}, Ramos {\it et al.} directly measured the tunneling time of Bose-condensed $^{87}$Rb atoms based on  Larmor clock technology for the first time \cite{nat15}. A Larmor clock uses an auxiliary degree of freedom (i.e., spin precession) of the tunneling atoms to measure the dwell time inside the 
barrier. Furthermore, they observed a slow decrease in the Larmor tunneling time with lower incident energy \cite{PRL2021}. 
%and emphasized that one should make a strict distinction between escape (or arrival) time and traversal (or interaction) time when discussing tunneling time. 
However, these studies limited the incident energy to a range greater than zero, while the intriguing quantum region where the incident energy is less than zero remains unsolved. 

In this Letter, we study the tunneling time for a wider energy range, especially for negative energies, which allows us to give a general description and to obtain a complete picture of the quantum tunneling time. To directly define the tunneling time without borrowing the auxiliary degrees of freedom of tunneling atoms, we consider a single-component BEC passing through a square barrier. We use the time difference between two moments when the probabilities of finding atoms at two boundaries are maximized to characterize the tunneling time. We explore the dependence of the tunneling time on the incident velocity (or energy) of the wave packet and find that the tunneling time exhibits different scaling laws in different energy regimes. Particularly, for negative incident energy, we see a counterintuitive phenomenon in which the tunneling time decreases rapidly with decreasing incident velocity and find that the tunneling time is maximal at a particular incident velocity when the incident energy is positive. We further investigate the effect of barrier width on the maximum tunneling time and find a critical barrier width that can be used to mark the boundary between classical and quantum regimes. The physical mechanism behind the results and their connection to wave--particle duality and uncertainty relation are discussed as well. Our results are qualitatively consistent with previous experimental measurements based on the Larmor clock and can be used to understand and explain the controversy over tunneling time in the field of strong-field ionization.

%We investigate the problem of tunneling time at the most basic level to understand the tunneling process itself more deeply and comprehensively. Moreover, our present research can establish links with existing Larmor clock experiments in two-component BEC and related research in the field of strong-field ionization.  
%Our intention is not to find a unique tunneling timescale, but to provide a physical definition that can be used to maximize understanding and accelerate existing experimental and theoretical studies of tunneling time.  

%reproduce the results of Larmor tunneling times for the atoms with higher incident energy. we focus on the tunneling of atoms with very low incident energy to get a more complete picture of tunneling time in the incident energy domain. For a square barrier with medium width, we divide the whole tunneling process into three distinct regions by comparing barrier height with incident energy. We find that the tunneling time in each region satisfies different scaling laws with incident velocity. 

A paradigmatic model for discussing the macroscopic tunneling dynamics of ultracold atoms is the one-dimensional Gross-Pitaevskii (GP) equation. For convenience, we adopt the dimensionless form, which reads   
\begin{align}\label{GP}
i\frac{\partial\psi(x,t)}{\partial t}=\left[-\frac{\partial^2}{2\partial x^2}+V(x)-u|\psi(x,t)|^2\right]\psi(x,t),
\end{align}
where $V(x)$ denotes a square potential barrier centered on $x=0$ with height $q$ and width $w$. The left and right boundaries of the barrier are located at $x_{L,R}=\pm\frac{w}{2}$. The atomic interaction constant $u$ is defined by
$u = 4\pi n_0l^2_0$, with $n_0$ being the maximum density in the initial distribution of the atomic condensates. The position
$x$, time $t$, and macroscopic wave function $\psi$ are, respectively in units of $l_0=1\mu$m, $ml_0^2/\hbar$, and $\sqrt{n_0}$, with $m$ being the atom mass and $\hbar$ being the reduced Planck constant. Initially, we assume that the matter-wave packet
takes the following normalized form:
\begin{equation}\label{ISECH}
\psi(x,t=0) = \frac{1}{\sqrt{2}}\mathrm{sech}(x+x_0)e^{ivx}.
\end{equation}
This expression approximates a matter-wave solution comprising one bright soliton with three parameters: $a_s$, $w_s$, and $v$, which denote the amplitude (i.e., ${1}/{\sqrt{2}}$), width (i.e., $2\mathrm{sech}^{-1}\frac{1}{2}\simeq 2.634$), and velocity, respectively.
The matter-wave soliton brings a good opportunity to investigate the wave--particle duality on a
macroscopic scale \cite{epl20} due to its particle-like qualities, such as localization, nondispersion \cite{epl15}, and center-of-mass trajectories \cite{epl17,epl18,epl19}. Soliton tunneling \cite{epl28,epl29,epl30,epl31} associated with the nonlinear dynamics of a wave packet colliding with a potential can be used to illustrate the link between classical and quantum mechanics \cite{epl20,eplme}. Initially, we set $x_0=-15$ (i.e., the initial position of the wave packet) to ensure that the wave packet is far enough away from the barrier. We take $u=2$ to keep the shape of the wave packet unchanged as it travels freely. In the calculations, we numerically solve the time-dependent GP equation via the time-splitting spectral method for the barrier height $q=2$. By comparing the incident velocity (or energy) of the wave packet with the barrier height, we can differentiate whether the motion of particles is in the classical regime or in the quantum regime.

%%%%%%%%%%%%%%%%%%%%%%%%%%%%%%%%%%%%%%%%%%%%%%%%%%%%%%%%%
\begin{figure}[t]
\centering
\includegraphics[width=0.75\linewidth]{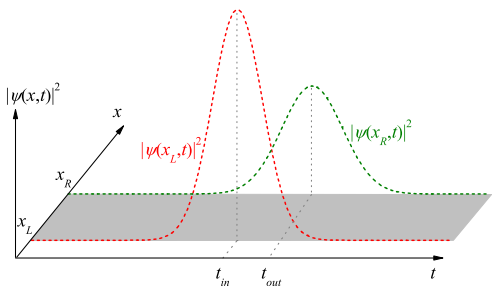}
\caption{Schematic of the quantification of tunneling time. We use $t_{in}$ and $t_{out}$ to mark the specific moments when particles enter and exit the barrier, respectively. The two moments correspond to the maximum probability of finding particles at the left and right boundaries of the barrier. The difference between them characterizes the tunneling time.}
\label{shiyifig}
\end{figure}
%%%%%%%%%%%%%%%%%%%%%%%%%%%%%%%%%%%%%%%%%%%%%%%%%%%%%%%%%%%%

Generally, to quantify a tunneling process, one must employ local quantities to clearly identify the start and end of tunneling events. A classical particle has a deterministic trajectory, and the time it takes to tunnel is uniquely determined as long as the position and momentum of the particle at the beginning and end are determined. A quantum particle is governed by the principle of uncertainty, and the position and momentum cannot be determined simultaneously. The motion of a quantum particle has no deterministic trajectory that can be traced; thus, there is no universal definition of the beginning and end of tunneling events. In our model (\ref{GP}), the microscopic state of a large number of condensed atoms is described by a macroscopic quantum local wave packet (\ref{ISECH}), and the trajectory of the collective motion of particles can be characterized by the center of mass. However, considering the deformation of the quantum wave packet when it collides with the barrier, we use the maximum probability at the barrier boundary to mark the beginning or the end of the tunneling event, as illustrated in Fig. \ref{shiyifig}. To obtain the tunneling time, we denote the start and end moments of the tunneling process by $t_{in}$ and $t_{out}$, respectively. These two moments are determined by the relations 
\begin{align}
&|\psi(x_L,t_{in})|^2=\max(|\psi(x_L,t)|^2),\\
&|\psi(x_R,t_{out})|^2=\max(|\psi(x_R,t)|^2).    
\end{align}
The difference between $t_{out}$ and $t_{in}$ gives the tunneling time, namely, $\Delta t$, 
\begin{equation}
\Delta t=t_{out}-t_{in},
\end{equation}
which characterizes the time spent by the atoms inside the barrier. 
%In this sense, our tunneling time is actually the interaction time or traversal time mentioned above rather than the so-called arrival time or escape time \cite{nat2020,PRL2021}.

\begin{figure}[t]
\centering
\includegraphics[width=0.9\linewidth]{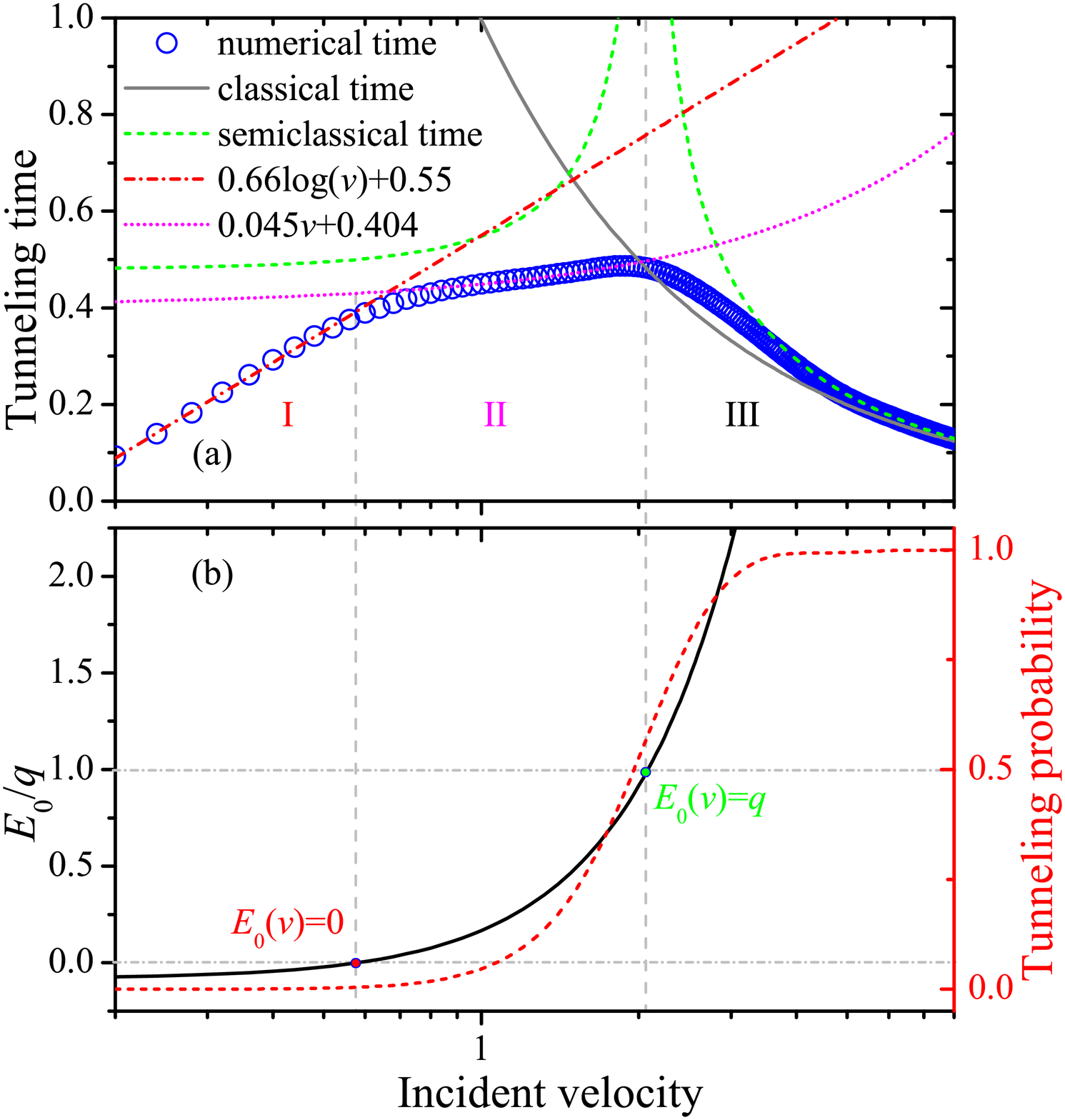}
\caption{(a) Tunneling times as functions of the incident velocity of the wave packet. The barrier width $w=1.0$ was used. The numerical data (hollow circle markers) along with the classical (solid line, $w/v$) and semiclassical (dashed lines, $w/\sqrt{2|q-E_0|}$) results are shown. The fits are illustrated by the dash-dotted and dotted lines, and their fitting parameters are reported in the text. The left vertical line represents the position where the initial energy of the wave packet equals zero [i.e., $E_0(v)=0$, red solid circle in (b)], and the right vertical line denotes the position where the initial energy matches the barrier height [i.e., $E_0(v)=q$, green solid circle in (b)]. (b) Initial energies (left axis) and tunneling probabilities (right axis) as functions of the incident velocity.}
\label{mainfig}
\end{figure}

The main point of this work is to study
the tunneling time in different energy regimes
and to determine the distinctive behaviors. Indeed, our system has two energy scales: the initial energy of the incident wave packet, i.e., $E_0=\int (\frac{1}{2}|\frac{\partial\psi}{\partial x}|^2-\frac{u}{2}|\psi|^4)dx$, and the energy corresponding to the barrier height, i.e., $q$. The initial energy includes two parts: the kinetic energy, i.e., $E_k$, and the potential energy, i.e., $E_p$. The kinetic energy consists of the intrinsic kinetic energy determined by the shape of the initial wave packet, i.e., $E_{ks}=E_{ks}(a_s,w_s)$, and the incident kinetic energy determined by $v$, i.e., $E_{kv}=E_{kv}(v)=\frac{1}{2}v^2$. For a given wave packet, the initial energy is a function of both the incident velocity of the wave packet and the strength of the interaction between atoms, i.e., $E_0=E_0(v,u)$. For Eq. (\ref{ISECH}), we have $E_0(v,u)=\frac{1}{6}(1 - u + 3 v^2)$, which contains $E_{ks}=\frac{1}{6}$, $E_{kv}=\frac{1}{2}v^2$, and  $E_p=-\frac{u}{6}$. When $u=2$, $E_0$ is simply a function of the incident velocity, i.e., $E_0(v)=\frac{1}{2}v^2-\frac{1}{6}$ or $v(E_0)=\sqrt{2E_0+1/3}$. Subsequently, we study the dependence of the tunneling time on the initial energy (or the incident velocity) of the wave packet, and the main results are shown in Fig. \ref{mainfig}. Comparing $E_0$ with $q$, we discuss the following three regimes separately.

(I) If $E_0(v)<0$ [see Fig. \ref{mainfig}(b)] or $v(E_0)<v(E_0=0)$, 
we find
\begin{equation}
\Delta t\simeq A\log(v)+B,    
\end{equation}
where the constants $A$ and $B$ are determined by the width of the barrier. For $w=1.0$, we obtain $A\simeq 0.66$ and $B\simeq 0.55$ [see Fig. \ref{mainfig}(a)]. In this region, the initial energy of the wave packet is negative. One can regard the wave packet as a quasibound state. The De Broglie wavelength is $\lambda_D={2\pi}/{v(E_0)}$, and its minimum value (i.e., ${2\pi}/{v(E_0=0)}=2\sqrt{3}\pi\simeq 10.883$) is larger than the width of the wave packet, i.e., $\lambda_D>w_s$. As a result, the deformation of the wave packet is very small when interacting with the barrier, and the tunneling probability is also very small [see Fig. \ref{mainfig}(b)]. However, the tunneling time in this region increases very rapidly with increasing  incidence velocity.
%%%%%%%%%%%%%%%%%%%%%%%%%%%%%%%%%%%%%%%%%%%%%%%%%%%%%%%%%%%%%%
%%%%%%%%%%%%%%%%%%%%%%%%%%%%%%%%%%%%%%%%%%%%%%%%%%%%%%%%%%%%%%

(II) If $0<E_0(v)<q$ or $v(E_0=0)<v(E_0)<v(E_0=q)$, we find
\begin{equation}
\Delta t\simeq \alpha v+\beta,    
\end{equation}
where the constants $\alpha$ and $\beta$ are determined by the width of the barrier. For $w=1.0$, we obtain $\alpha\simeq 0.045$ and $\beta\simeq 0.404$ [see Fig. \ref{mainfig}(a)]. In this region, the initial energy of the wave packet is positive but less than the height of the barrier. The De Broglie wavelength is between ${2\pi}/{v(E_0=q)}={2\pi}/\sqrt{2q+1/3}\simeq 3.078$ and ${2\pi}/{v(E_0=0)}=2\sqrt{3}\pi$ and is comparable to the wave packet width, i.e., $\lambda_D\sim w_s$. Before the wave packet arrives at the left boundary of the barrier, we find that the deformation of the packet is remarkable and that the tunneling probability significantly increases [see Fig. \ref{mainfig}(b)]. However, the tunneling time in this region is proportional to the incidence velocity and changes very slowly. In particular, we find that the tunneling time is remarkably shorter than that predicted by semiclassical theory, i.e., $w/\sqrt{2(q-E_0)}$.

(III) If $E_0(v)>q$ [see Fig. \ref{mainfig}(b)] or $v(E_0)>v(E_0=q)$, 
we find 
\begin{equation}
\frac{w}{v}\leq\Delta t\leq \frac{w}{\sqrt{2(E_0-q)}}.    
\end{equation}
In this regime, the tunneling time is more than the classical time $w/v$ but less than the semiclassical time $w/\sqrt{2(E_0-q)}$ [see Fig. \ref{mainfig}(a)]. The De Broglie wavelength is less than  ${2\pi}/{v(E_0=q)}={2\pi}/\sqrt{2q+1/3}$, which quickly decreases and is shorter than the width of the wave packet, i.e., $\lambda_D< w_s$. It is noted that the tunneling probability in this region is greater than 50\%  [see Fig. \ref{mainfig}(b)]. When $v^2\gg q$ or $E_0\gg q$, the difference between the classical and semiclassical times disappears, and our numerical results tend to the theoretical predictions [see Fig. \ref{mainfig}(a)]. 

From the perspective of wave--particle duality, the above three regions show the situations  where waves are dominant, waves and particles are comparable, and particles are dominant, respectively. Thus, we can naturally regard I as the quantum regime, II as the semiclassical regime, and III as the classical regime [see Fig. \ref{mainfig}]. It is clear that our theoretical calculation of tunneling time is qualitatively consistent with the recent experimental results of Larmor tunneling times in the semiclassical and classical regimes \cite{nat2020,PRL2021}. A maximum tunneling time $\Delta t_{\max}$ at $v\simeq\sqrt{2q}$, where the total kinetic energy of the wave packet matches the barrier height, is also found. Surprisingly, our new results in the quantum regime seem counterintuitive but do support some of the findings in strong-field tunneling ionization \cite{nat7}. That is, in the quantum limit $v\rightarrow 0$, the tunneling of microscopic particles seems instantaneous.

\begin{figure}[t]
\centering
\includegraphics[width=0.9\linewidth]{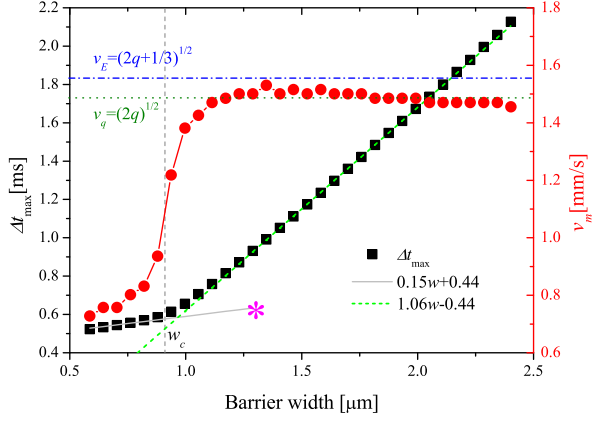}
\caption{The maximum tunneling time (left axis, black solid squares) and the corresponding incident velocity of the wave packet (right axis, red solid circles) as functions of the barrier width. The pink asterisk marks the experimental result given in Ref. \cite{nat2020}. Two horizontal lines mark the velocities, which are determined by $E_0(v)=q$ ($v_E$, olive dash-dotted line) and $\frac{1}{2}v^2=q$ ($v_q$, blue dotted line). The vertical gray dotted line represents the critical barrier width $w_c$, which denotes the boundary between the quantum and classical regions.}
\label{mainfiga}
\end{figure}

The above results for fixed barrier height and barrier width are dimensionless. In practical measurements, one should pay more attention to the dimensional maximum tunneling time. For example, one can use $^7$Li atomic condensate to generate the wave packet \cite{np2014}. For the barrier with height $q=2\mu$m and width $w=1\mu$m, when the incident velocity of the wave packet is $v=0.017$mm/s, our theoretical prediction for the maximum tunneling time is $\Delta t_{\max}=0.053$ms. If one uses $^{87}$Rb atoms, when the incident velocity of the wave packet is $v=1.382$mm/s, our prediction becomes $\Delta t_{\max}=0.656$ms, which is of the same order as the experimental result [i.e., $0.61(7)$ms, we have marked this point with an asterisk in Fig. \ref{mainfiga}.] based on the Larmor clock \cite{nat2020}. The difference between the two results is mainly caused by the difference in both the shape of the barrier and the incident velocity. Actually, both the maximum tunneling time and the corresponding wave packet incidence velocity depend on the barrier parameters and the shape of the incident wave packet, i.e., $\Delta t_{\max}=\Delta t_{\max}(q,w;a_s,w_s)$ and $v_m(q,w;a_s,w_s)$. For our given wave packet, they are functions of the width and height of the barrier, i.e., $\Delta t_{\max}=\Delta t_{\max}(q,w)$ and $v_m=v_m(q,w)$.
%%%%%%%%%%%%%%%%%%%%%%%%%%%%%%%%%%%%%%%%%%%%%%%%%%%%%%%%%%%%
%%%%%%%%%%%%%%%%%%%%%%%%%%%%%%%%%%%%%%%%%%%%%%%%%%%%%%%%%%%%%%%

 To gain insight into the connection between $\Delta t_{\max}$ and $v_m$, we discuss the influence of $w$ and for a fixed barrier height $q$. Figure \ref{mainfiga} shows the variation of both $\Delta t_{\max}$ and $v_m$ with $w$ for $^{87}$Rb atoms. We mark the width where $v_m$ changes most dramatically as $w_c$ which approximately equals $0.889\mu$m. When $w<w_c$, we find that $\Delta t_{\max}$ increases slowly with $w$ as $\Delta t_{\max}\simeq (0.15w+0.44)$ms. When $w>w_c$, we see that $\Delta t_{\max}$ increases rapidly with $w$ as $\Delta t_{\max}\simeq (1.06w-0.44)$ms. It is clear that these two different kinds of changes in $\Delta t_{\max}$ follow good linear rules. Unlike the gradual change in the maximum tunneling time (see solid squares in Fig. \ref{mainfiga}), the corresponding incident velocity dramatically increases near the critical barrier width and then reaches a saturation value (see solid circles in Fig. \ref{mainfiga}). It should be mentioned that the saturation velocity is less than the critical velocity $v_E$ (i.e., $E_0(v)=q$, horizontal blue dash-dotted line) which indicates the boundary between the semiclassical and classical regions and is approximately equal to the velocity $v_q$ (i.e., $v=\sqrt{2q}$, horizontal olive dotted line). The critical barrier width $w_c$ corresponds to the position ($\Delta t_{\max}^{\ast}\simeq0.594$ms, $v_m^{\ast}\simeq 1.068$mm$/$s) where the wave packet kinetic energy changes most dramatically. In other words, one can use the property that $v_m$ is sensitive to $w$ to  identify $w_c$.

We can analyze the above results by using the uncertain relation between energy and time. With the help of $v_E=\sqrt{2q+{1}/{3}}\simeq 1.546$mm$/$s, we have the maximum energy difference $\Delta E_{\max}=\frac{1}{2}m(v_E^2-v_m^2)$. When $w<w_c$, $\Delta E_{\max}\cdot\Delta t_{\max}>\Delta E_{\max}^{\ast}\cdot\Delta t_{\max}^{\ast}\simeq\frac{\hbar}{2}$ with $\Delta E_{\max}^{\ast}=\frac{1}{2}m[v_E^2-(v_m^{\ast})^2]$. In this regime, the uncertainty relation is satisfied,  and thus, the tunneling particles have pure quantum properties. When $w>w_c$, $\Delta E_{\max}\cdot\Delta t_{\max}<\Delta E_{\max}^{\ast}\cdot\Delta t_{\max}^{\ast}\simeq\frac{\hbar}{2}$. In this region, the uncertainty relation is broken down, and therefore, the tunneling particles have obvious classical properties. We can also analyze the above results by using the uncertain relation between momentum and space. We use $\Delta p_{\max}$ and $\Delta x_{\max}$ to record the maximum uncertainty ranges of momentum and position, respectively, which can be expressed as $\Delta p_{\max}=m(v_q-v_m)$ and $\Delta x_{\max}=w$ with $v_q\simeq 1.486$mm$/$s. Similarly, we find that when $w<w_c$, the uncertain relation $\Delta p_{\max}\cdot\Delta x_{\max}>\Delta p_{\max}^\ast\cdot\Delta x_{\max}^\ast\simeq\frac{\hbar}{2}$ holds with $\Delta p_{\max}^\ast=m(v_q-v_m^\ast)$ and $\Delta x_{\max}^\ast=w_c$. When $w>w_c$, the uncertain relation $\Delta p_{\max}\cdot\Delta x_{\max}>\frac{\hbar}{2}$ is violated. This implies that when discussing quantum tunneling, both energy requirements (i.e., $E_0<q$ or $E_0<0$) and barrier width constraints (i.e., $w<w_c=\frac{\hbar}{2m(v_q-v_m)}$) should be considered, because  they together determine the boundary  between quantum and classical regimes.

In summary, we have presented a general
and detailed study of the tunneling time by investigating the collisions between
a macroscopic quantum wave packet and a square barrier. We explore the dependence of the tunneling time on the incident energy of the wave packet and find three different time scales in different energy regimes. In particular, the counterintuitive behavior that the tunneling time increases rapidly with the incidence velocity is shown in the negative incident energy regime, where the tunneling probability is very small. We also investigate the effect of the barrier width on the maximum tunneling time and find a critical barrier width, which together with the incident energy requirement determines the boundary between the quantum and classical regimes. We demonstrate the important role of both wave--particle duality and the uncertainty principle in the study of tunneling time. 
Our results are not only qualitatively consistent with the Larmor clock experiments but also support the claim of instantaneous tunneling in the study of strong-field ionization. Our study shows a comprehensive physical picture for understanding and studying the tunneling time and provides a new perspective to resolve the controversy over the tunneling time in strong-field ionization.

%\section*{Acknowledgement}
This work is supported by the National Natural Science Foundation of China (Grant No. 11974273).

%\bibliography{references.bib}

%

\end{document}